\documentstyle[12pt,aasms4,flushrt]{article}

\received{}
\accepted{}

\begin{document}

\title{86, 43, and 22 GHz VLBI Observations of 3C~120}

\author{Jos\'e-Luis G\'omez} 
\affil{Instituto de Astrof\'{\i}sica de Andaluc\'{\i}a, CSIC, Apartado 3004,
18080 Granada, Spain}
\authoremail{jlgomez@iaa.es}

\author{Alan P. Marscher}
\affil{Department of Astronomy, Boston University, 725 Commonwealth
Ave., Boston, MA 02215, USA}
\authoremail{marscher@bu.edu}
\and
\author{Antonio Alberdi}
\affil{Instituto de Astrof\'{\i}sica de Andaluc\'{\i}a, CSIC,
Apartado 3004, 18080 Granada, Spain}
\authoremail{antxon@iaa.es}

\begin{abstract}

  We present the first 86 GHz VLBI observations of the radio galaxy 3C~120,
together with contemporaneous 43 and 22 GHz polarimetric VLBA observations.
The very high angular resolution obtained at 86 GHz provides an upper limit to
the size of the core of 54 $\mu$as (0.025 $h^{-1}$pc). This represents a
direct determination of the base of the jet which is independent of
variability arguments (which depend on uncertain estimates of the Doppler
factor), and places it below approximately one light-month. Comparison with
previous VLBA observations after a one-year interval shows pronounced changes
in the structure and polarization of the jet. Most of the components are found
to follow a curved path while undergoing a steepening of their spectra
accompanied by a decrease in total and polarized emission. However, at least
one component is observed to follow a quasi--ballistic motion, accompanied by
a flattening of its spectrum, as well as an increase in total and polarized
flux. This may be explained by its interaction with the external medium,
resulting in a shock that enhances the emission and aligns the magnetic field
perpendicular to the component motion, thereby producing an increase of the
degree of polarization from undetected values to as high as 15\%. A second
strong component, with the highest degree of polarization (23\%), is found to
have experienced a displacement from the ridge line of the structural position
angle of the jet as it moved downstream. We have found a mean swing to the
south of the position angle of the innermost components of $\sim 6^{\circ}$
between late 1996 and 1997, which may be responsible for the jet curvature
observed at parsec and kiloparsec scales.


\end{abstract}

\section{Introduction}

  The radio galaxy 3C~120 is a powerful and variable emitter of radiation at
all observing frequencies. It is usually classified as a Seyfert 1 galaxy
(Burbidge \cite{B67}), although its optical morphology is also consistent with
that of a broad-line radio galaxy. It was among the first sources in which
superluminal motion was detected, on a scale of parsecs (Seielstad et
al. \cite{Se79}; Walker, Benson, \& Unwin \cite{Wa87}) to tens of parsecs
(Benson et al. \cite{Be88}; Walker \cite{Wa97}). An optical counterpart of the
radio jet has also been detected (Hjorth et al. \cite{Hj95}). 3C~120 is among
the closest known extragalactic superluminal sources ($z$=0.033), allowing the
study of its inner jet structure with unusually fine linear resolution for
this class of objects. Previous observations using NRAO's\footnote{The
National Radio Astronomy Observatory is a facility of the National Science
Foundation operated under cooperative agreement by Associated Universities,
Inc.} Very Long Baseline Array (VLBA) at 22 and 43 GHz (G\'omez et
al. \cite{JL98}, hereafter G98) revealed a very rich inner jet structure
containing up to ten different superluminal components, with velocities
between 2.3 and 5.5 $h^{-1}c$, mapped with a linear resolution of 0.07
$h^{-1}$pc. Linear polarization was also detected in several components,
revealing a magnetic field orientation that varies with respect to the jet
flow direction as a function of frequency, epoch, and position along the jet.

  In this {\it Letter} we present the first 86 GHz Coordinated Millimeter VLBI
Array (CMVA) observations of 3C~120, providing an angular resolution of 54
$\mu$as, which for this source represents a linear resolution of 0.025
$h^{-1}$pc. We also present contemporaneous 22 and 43 GHz polarimetric VLBA
observations of 3C~120 and compare these with previous observations obtained
one year previously.

\section{Observations and data analysis}

  The 86 GHz CMVA observation took place on 1997 October 24 (epoch
1997.82). Participating antennas were Pico Veleta, Effelsberg, Onsala,
Mets\"ahovi, Haystack, Kitt Peak 12-Meter, and VLBA Pie Town. The data were
correlated with the Haystack MkIII correlator, after which global fringe
fitting was performed within the NRAO Astronomical Image Processing System
(AIPS) software. Because of bad weather at some of the stations and the
relatively low flux density exhibited by 3C~120 during the observations, a
significant amount of data were lost. We found a very consistent calibration
for the American sub-array of antennas, which was used to calibrate the rest.

  The 7 mm and 1.3 cm VLBA observations were performed on 1997 November 10,
seventeen days apart from the CMVA observation. The data were recorded in
1-bit sampling VLBA format with 32 MHz bandwidth per circular
polarization. The reduction of the data was performed with the AIPS software
in the usual manner (e.g., Lepp\"anen, Zensus, \& Diamond
\cite{Ka95}). Opacity corrections were introduced by solving for receiver
temperature and zenith opacity at each antenna. Fringe fitting to determine
the residual delays and fringe rates was performed for both parallel hands
independently and referred to a common reference antenna. Delay differences
between the right- and left-handed polarization systems were estimated over a
short scan of cross-polarized data of a strong calibrator (3C~454.3). The
instrumental polarization was determined by using the feed solution algorithm
developed by Lepp\"anen, Zensus, \& Diamond (\cite{Ka95}).

  The absolute phase offset between right and left circular polarization at
the reference antenna was determined by VLA observations of the sources
0420-014 and OJ~287 on 1997 November 21, and referenced to an assumed
polarization position angle of 33$^{\circ}$ for 3C~286 and 11$^{\circ}$ for
3C~138, at both observing frequencies. This provided an estimation of the
absolute polarization position angle within 9$^{\circ}$ and 7$^{\circ}$ at 22
and 43 GHz, respectively.

\section{Results and conclusions}

  Figure 1 shows the resulting CMVA and VLBA images of 3C~120. Table 1
summarizes the physical parameters obtained for 3C~120 at the three
frequencies. Tabulated data correspond to total flux density ($S$), polarized
flux density ($P$), magnetic vector position angle ($\chi_B$), separation
($r$) and structural position angle ($\theta$) relative to the easternmost
bright component [which we refer to as the ``core''], and angular size
(FWHM). Components in the total intensity images were analyzed by model
fitting the uv data with circular Gaussian components within the software
Difmap (Shepherd \cite{Se97}). We have obtained an estimation of the errors in
the model fitting by introducing small changes in the fitted parameters and
studying the variations in the $\chi^2$ of the fit, as well as reproducing the
model fitting from the start and comparing with previous results. For the 43
GHz image the estimated model fitting errors are of the order of 5 mJy in
flux, between 0.01 and 0.02 mas in position, and $\sim$ 0.01 in component
sizes. Larger errors, of the order of 50\% higher, are estimated for the 22
GHz image model fitting.

  The 3 mm CMVA image of Fig. 1 shows the existence of a bright core
component. The mapping process was able to recover 1.51 Jy of the integrated
total flux of 1.95$\pm$0.05 Jy measured by Pico Veleta (H. Ungerechts, private
communication), resulting in a maximum brightness temperature of
1.15$\times$10$^{10}$ K. The combined spectrum for the core at the three
observing frequencies gives an inverted spectrum with $\alpha \sim$ 1.1
($S_{\nu} \propto \nu^{\alpha}$). Components n1 and n2 show a steep spectrum
between 22 and 43 GHz, which, together with the inverted spectrum of the core,
would explain why they are not detected in the 86 GHz map, given the
relatively poor dynamic range of the latter. Mets\"ahovi flux density
monitoring at 22 and 37 GHz (H. Ter\"asranta, private communication) observed
the strongest radio flare ever seen in 3C~120 to take place starting in the
end of 1997 and peaking in June 1998, which would explain the flattening of
the spectrum of the core with respect to the late 1996 observations. No
polarization was detected for the core at either 22 or 43 GHz.

  We estimate the size of the core to lie below the resolution of the 86 GHz
map, that is $\leq$ 54 $\mu$as, which implies a linear size less than 0.025
$h^{-1}$pc, or 30 light-days. This represents a direct determination of the
upper limit to the size of the base of the jet that is independent of
variability arguments, which depend on the quite uncertain estimate of the
Doppler factor.

  The 43 and 22 GHz images show a much richer structure consisting of multiple
components. We have labeled components from west to east, using upper-case
letters for the map at 22 GHz. We shall note that several of the components in
the 43 GHz image appear blended in the corresponding 22 GHz image. Through
comparison with previous images obtained at the end of 1996 (G98) and
extrapolation of the component speeds, we have cross--identified several
components. The most reliable identification corresponds to components A, B,
C, and D. Components H and K may be associated with two distinct radio flares
detected by Mets\"ahovi (H. Ter\"asranta private communication) at the end of
1996 and in mid-1997, with component H appearing previously in the November
1996 and December 1996 images of G98.

  In comparison with the late 1996 epochs of G98, components A, B, C, E, F,
and G are found to have experienced a similar evolution. Each moved along the
curved path traced by the jet in the plane of the sky. During this downstream
translation their flux densities declined at both observing frequencies
(except for component E at 43 GHz). This is as expected for components that
are subject to expansion and radiative cooling (the latter of which might have
occurred earlier and become prominent at radio frequencies only after further
expansion). There is some evidence for deceleration of components upstream of
D to a common value close to $\sim 3 \, h^{-1}c$, but further observations are
needed to confirm this. One of the largest variations can be found in
polarization. Components F and G decreased their polarized flux density
between November and December 1996 to undetectable values, but by November
1997 showed an increase to values between 9 and 14\%. This change was
accompanied by a rotation of the magnetic field vector from orthogonal to
parallel to the jet axis. Components B and C experienced a similar drastic
variation, changing their degree of polarization from values close to 10\% in
late 1996, to values below the noise in the images in Fig. 1. This suggests
changes in the underlying configuration of the magnetic field and/or the
shocks presumably associated with the components in the inner parsec of
3C~120.

  We found a significantly different evolution for component D. During the
one-year period between the G98 observations and that shown in Fig. 1,
component D gained flux, and its spectrum became significantly flatter. Even
more pronounced changes are found in polarization, from below the noise level
in the G98 observations, to a level that made it one of the strongest
components in polarized flux density, with a degree of polarization of 15\%
(13\% at 43 GHz). All this evidence suggests significant activity in component
D not present in the other components.

  We interpret the different evolution occurring in component D as caused by
its interaction with the external medium. As shown in the 22 GHz image, the
jet clearly bends at the position of D, hence we expect at least a standing
oblique shock there. While between late 1996 and November 1997 the position
angle of components B and C rotated by $\sim$ 3$^{\circ}$, the increment in
component D was significantly smaller, of the order of 1.5$^{\circ}$. As a
consequence, the curvature near component D appears more pronounced than in
the previous images of G98, actually showing component D centered $\sim$ 0.3
mas south of the jet axis defined by components E and C. If the jet components
travel along a gently curved jet funnel, it is possible that components with
relatively larger momentum --as may be the case for D-- would move along a
more ballistic trajectory, progressively moving closer to the jet funnel
boundary, and consequently undergoing interactions with the external
medium. This would explain the more ballistic trajectory of component D, while
other components seem to have followed the curvature of the jet. The
interaction with the external medium should produce a strong shock and
turbulence (from Rayleigh-Taylor and perhaps Kelvin-Helmholtz instabilities),
enhancing the emission and flattening the spectrum, as is commonly observed in
the terminal hot spot of large-scale jets. The shock would also produce an
enhancement and reordering of the magnetic field parallel to the shock front;
this would explain the increase in the polarized flux and degree of
polarization. The magnetic field orientation in D is consistent with this
scenario, and contrasts with that observed for the rest of the jet, which is
roughly parallel to the jet flow (except for components n1 and M).

  Component H seems to share some of the properties and activity of component
D. Its structural position angle shifted to the south by 2.5$^{\circ}$, which
placed it $\sim$ 0.25 mas south of the jet axis in Nov 1997. The spectrum is
flatter at this later epoch, and the degree of polarization increased, from
values below 4\% at the end of 1996, to 23\% (18\% at 43 GHz) in Nov 1997, the
maximum value measured in the jet. During this burst in polarization, H
maintained a similar orientation of its magnetic field, almost parallel to the
jet axis, with a slightly shifted orientation at 43 GHz towards the direction
of component d's field. To the north of H (h at 43 GHz), model fitting reveals
the presence of component I (i), which presumably is traveling through the jet
funnel, out of which H seems to be breaking. This is suggested by a slightly
different orientation of the magnetic field position angle.

  The structural position angle of the inner milliarcsecond components in
3C~120 underwent a mean shift to the south of $\sim 6^{\circ}$ between late
1996 and 1997. This indicates a progressive swing of the jet ejection
direction towards the south, which may be responsible for the continuous
curvature observed on parsec and kiloparsec scales (e.g., Walker
\cite{Wa97}). It is then expected that some components (perhaps with
relatively larger momentum) --- such as D and H --- would experience strong
interactions with the external medium, opening a new path through which
subsequently ejected components would travel relatively undisturbed.
  
  Further observations (in progress) will provide a follow up of the changes
observed in 3C~120 and test the hypothesis presented here, as well as allow
comparison with simulations of the hydrodynamics and emission of relativistic
jets (G\'omez et al. \cite{JL97}), resulting in a better understanding of the
physics involved in 3C~120 and other AGNs.

\begin{acknowledgements}
This research was supported in part by Spain's Direcci\'on General de
investigaci\'on Cient\'{\i}fica y T\'ecnica (DGICYT), grants PB94-1275 and
PB97-1164, by NATO travel grant SA.5-2-03 (CRG/961228), and by U.S. National
Science Foundation grant AST-9802941. We thank Iv\'an Agudo-Rodr\'{\i}guez for
preparing Fig. 1. We thank Robert Phillips and Elisabeth Ambrose for providing
confirmation of the CMVA results through a map obtained using the software
HOPS. We are also grateful to Barry Clark for providing \emph{ad hoc} VLA time
in order to determine the absolute polarization position angle.
\end{acknowledgements}

\newpage
\begin{figure*}[t]
\figcaption{VLBI images of 3C~120 at 86 ({\it top}), 43
({\it middle}), and 22 GHz ({\it bottom}). Total intensity is plotted as
contour maps, while the linear gray scale shows the linearly polarized
intensity (22 and 43 GHz only).  The superposed sticks give the direction of
the {\it magnetic field} vector. For the 86, 43, and 22 GHz images
respectively: Contour levels are in factors of 2, starting at (noise level) 2,
0.9, and 0.5\% of the peak intensity of 1.51, 0.54, and 0.48 Jy beam$^{-1}$
(43 and 86 GHz images contain an extra contour at 90\% of the
peak). Convolving beams (shown to the lower left of the core of each image)
are 0.40$\times$0.054, 0.32$\times$0.16, and 0.63$\times$0.30 mas, with
position angles of -13$^{\circ}$, -6$^{\circ}$, and -4$^{\circ}$. Peaks in
polarized intensity are 27, and 42 mJy beam$^{-1}$, with noise levels of 3 and
2 mJy beam$^{-1}$ for the 43 and 22 GHz images, respectively. Epoch and scale
of each map are shown on the right (note that 86 GHz map scale is twice of the
other maps).}
\end{figure*} 

\newpage

\begin{tiny}
\begin{deluxetable}{llcccccccccccc}
\tablecolumns{14}
\tablewidth{0pc}
\tablecaption{Physical Parameters of 3C~120 \label{tab1}}
\tablehead{
\multicolumn{2}{c}{COMPONENTS}&
\multicolumn{2}{c}{$S$}&
\multicolumn{2}{c}{$P$}&
\multicolumn{2}{c}{$\chi_B$}&
\multicolumn{2}{c}{$r$}&
\multicolumn{2}{c}{$\theta$}&
\multicolumn{2}{c}{$FWHM$}\nl
 & &
\multicolumn{2}{c}{(mJy)}&
\multicolumn{2}{c}{(mJy)}&
\multicolumn{2}{c}{($^{\circ}$)}&
\multicolumn{2}{c}{(mas)}&
\multicolumn{2}{c}{($^{\circ}$)}&
\multicolumn{2}{c}{(mas)}}
\startdata
13mm&7mm&13mm&7mm&13mm&7mm&13mm&7mm&13mm&7mm&13mm&7mm&13mm&7mm\nl
\hline
\multicolumn{2}{c}{Core$^a$\dotfill}  
                 &324&551&...   &...   &...   &...   &... &... &...   &...   &0.11   &0.04\nl
N &n2   \dotfill &374&109&...   &...   &...   &...   &0.27&0.21&-132.4&-130.2&$<$0.01&0.01\nl
  &n1   \dotfill &...&144&...   &5     &...   &100   &... &0.34&...   &-131.3& ...   &0.06\nl
M &m    \dotfill &129& 50&4     &...   &113   &...   &0.61&0.52&-131.9&-130.3&$<$0.01&0.10\nl
L &l    \dotfill &205& 71&...   &14    &...   &59    &0.95&0.84&-130.7&-133.1&0.19   &0.16\nl
K &k2   \dotfill &398&193&12    &21$^b$&55    &43$^b$&1.17&1.11&-122.6&-125.6&0.20   &0.22\nl
  &k1   \dotfill &...&173&...   &21$^b$&...   &43$^b$&... &1.28&...   &-122.0& ...   &0.15\nl
J &j    \dotfill & 64& 25&4     &4     &34    &34    &1.60&1.58&-121.6&-121.5&0.22   &0.14\nl
  &i2   \dotfill &...& 92&...   &...   &...   &...   &... &2.09&...   &-120.7&...    &0.31\nl
I &i    \dotfill &257& 37&50$^c$&33$^d$&69$^c$&70$^d$&2.17&2.29&-120.9&-122.0&0.31   &0.05\nl
H &h    \dotfill &213&183&50$^c$&33$^d$&69$^c$&80$^d$&2.28&2.33&-127.2&-126.8&$<$0.01&0.02\nl
G2&g2   \dotfill & 15& 26&...   &...   &...   &...   &2.53&2.44&-118.2&-119.2&0.16   &0.17\nl
G &g    \dotfill & 96& 58&13    &8     &65    &41    &3.07&3.12&-113.8&-113.4&0.40   &0.38\nl
F &f2   \dotfill &117& 30&11$^e$&...   &64$^e$&...   &3.64&3.60&-114.7&-114.2&0.48   &0.24\nl
  &f1   \dotfill &...& 40&...   &4     &...   &47    &... &3.94&...   &-114.8& ...   &0.35\nl
E &e    \dotfill & 38& 14&...   &...   &...   &...   &4.42&4.60&-114.1&-113.7&0.50   &0.14\nl
D &d    \dotfill &156&124&23    &15    &125   &118   &5.26&5.32&-115.7&-115.8&0.15   &0.17\nl
C &     \dotfill & 45&...&...   &...   &...   &...   &5.59&... &-108.3&...   &0.76   &...\nl
B &     \dotfill & 52&...&...   &...   &...   &...   &6.34&... &-104.0&...   &0.93   &...\nl
A &     \dotfill & 29&...&...   &...   &...   &...   &7.90&... &-107.3&...   &1.19   &...\nl
\enddata

\tablenotetext{a}{CMVA 3mm observations reveal a total flux density of 1.51
Jy. Pico Veleta obtained an integrated total flux density of 1.95$\pm$0.05
Jy (H. Ungerechts, private communication).}
\tablenotetext{b}{Common value for components k1 and k2, both blended in polarization.}
\tablenotetext{c}{Common value for components I and H, both blended in
polarization.}
\tablenotetext{d}{Although components i and h appear blended in polarized
flux, a slightly different $\chi_B$ can be observed.}
\tablenotetext{e}{There is some extension in polarization northwest of F with
a flux density of 8 mJy and $\chi$=93$^{\circ}$.}

\end{deluxetable}
\end{tiny}

\end{document}